# Quantum Čerenkov Radiation: Spectral Cutoffs and the Role of Spin and Orbital Angular Momentum

Ido Kaminer[1], Maor Mutzafi[2], Amir Levy[1], Gal Harari[2],

Hanan Herzig Sheinfux[2], Scott Skirlo[1], Jonathan Nemirovsky[2],

John D. Joannopoulos[1], Mordechai Segev[2], and Marin Soljačić[1]

1 - Department of Physics, Massachusetts Institute of Technology, Cambridge, MA 02139, USA
2 – Physics Department, Technion, Israel Institute of Technology, Haifa 32000, Israel

## Abstract

We show that the well-known Čerenkov Effect contains new phenomena arising from the quantum nature of charged particles. The Čerenkov transition amplitudes allow coupling between the charged particle and the emitted photon through their orbital angular momentum (OAM) and spin, by scattering into preferred angles and polarizations. Importantly, the spectral response reveals a discontinuity immediately below a frequency cutoff that can occur in the optical region. Specifically, with proper shaping of electron beams (ebeams), we predict that the traditional Čerenkov radiation angle splits into two distinctive cones of photonic shockwaves. One of the shockwaves can move along a backward cone, otherwise considered impossible for Čerenkov radiation in ordinary matter. Our findings are observable for ebeams with realistic parameters, offering new applications including novel quantum optics sources, and open a new realm for Čerenkov detectors involving the spin and orbital angular momentum of charged particles.

When a charged particle travels faster than the phase velocity of light in a medium, it produces Čerenkov radiation (ČR). The ČR ordinarily has a very broad spectrum, with its intensity increasing as the photon energy, thus famously seen as a "bluish glow". In principle, the spectrum extends up to a cutoff at the energy of the charged particle, though it is more commonly cut much earlier by material dispersion. Although it is now 80 years since its first observation [1], surprisingly little attention was given to the importance of the quantum nature of the charged particles producing the radiation. Since its discovery, the Čerenkov Effect has become a fundamental part of many fields [2]: devices like the Ring-imaging Čerenkov detector are used for cosmic radiation measurements [3,4]; it is serving as the basis of novel acceleration methods [5], and even as an unusual imaging tool in biology [6,7]. Due to the fundamental nature of ČR, it is found in many different physical systems, such as in nonlinear optics 8-11], used in the design of quantum cascade lasers [12], and predicted to yield the generation of entangled photon pairs [13-14]. Other kinds of ČR were found in photonic crystals [15-16], tunable light sources [17], coherently driven ultracold atomic gas [18], and recently even in active gain medium [19]. Many more novel ČR effects are still being found in new settings, such as surface polaritons [20] and metamaterials [21], where recent findings are suggesting to revolutionize Čerenkov detectors [22]. Even nanoparticles are now being combined with the Čerenkov Effect in the UV radiation from charges emitted from radioactive isotopes, allowing in-depth phototherapy [23]. However, despite the immense progress and the many generalizations, researchers still regard the original ČR as the electromagnetic field emitted by a ***point charge*** moving with a relativistic speed. This is the exact same theory (henceforth referred to as the conventional theory) developed in 1937 by Frank and Tam [24], who later shared the Nobel Prize in Physics of 1958 with Čerenkov, for explaining his observation.

In fact, the studies that did address the Čerenkov Effect in a quantum mechanical formalism, have found an excellent correspondence to the conventional theory. Notably, Ginzburg was the first to work on the quantum ČR [25] already in 1940, during his PhD, by considering an electron as a single momentum state (a plane-wave). He stated that the quantum result *"coincides with the classical expression with accuracy up to infinitely small terms"*. In 1948 [26] ČR was re-derived by considering a "plane-wave electron" interacting with a photon through the Dirac Hamiltonian. These works showed that the only correction to the Frank-Tamm formula occurs for extremely energetic photons, with energy close to the rest mass of the electron. Ginzburg himself noted [25] that *"The quantum condition for radiation is different from the classical condition, but practically coincides with it for radiation in the visible and ultra-violet regions, in which we are interested"*. Other early works considered Hamiltonians that also quantized the medium, in order to study the implications of materials' resonances, as corrections to the homogenized permittivity in the conventional ČR (e.g. [27]). A 1957 review paper described the ČR from particles with different spins [28]. Another 1957 paper [29] studied how two-photon emission affects ČR, with later work providing observation of such corrections arising from higher order Feynman diagrams [30], and effects of external magnetic fields [31]. However, to the best of our knowledge, all previous research always reconfirmed the same classical limit found by Čerenkov, Frank and Tamm [1,24]. In fact, no previous work has ever predicted that the quantum properties of the particle may cause significant deviation from the conventional ČR in low photon energies, such as UV, optical, or below.

Here, we study the Čerenkov radiation emitted from charged particles described by a quantum wavepacket. We show that an upper frequency cutoff can be brought down to the optical region, above which the ČR is zero. Immediately below this cutoff, we find a discontinuity in the radiation spectrum, presenting a clear deviation from the conventional ČR theory that displays no such cutoffs or discontinuities whatsoever. We develop the selection rules coupling spin and OAM of the wavepacket representing the charged particle with the polarization and OAM of the emitted ČR, showing that these relations cause preferential emission angles creating a complex final state of quantum correlated electron-photon. We find exact closed-form expressions for the photon emission rate, showing that the resonance emission is sensitive to the change of the incoming particle spin (flip/non-flip) and to the polarization of the emitted photon. This implies that certain spin flip transitions can dominate the process, even when the emitted photon is in the optical range, created from a modestly relativistic electron. Importantly, we show that the effects are not specific for a particular choice of incident wavepackets, but persist for general relativistic cylindrical wavepackets. Some of the predictions (such as the rates of ČR emission) are also relevant beyond quantum wavepackets, also occurring for incoherent wavepackets, i.e., they only depend on the shape of the ebeam. In such cases, the ebeam is a classical superposition of charged particles, whose structure affects the rate of the emitted ČR. As such, our predictions should occur in many experimental scenarios, without special engineering of the particle wavepacket. These new effects can lead to surprising applications such as enhanced Čerenkov detectors, new sources of monochromatic radiation, and novel mechanisms for generation of quantum states.

The effects of specific wavepackets of the charged particle on ČR are related to fundamental questions arising from the quantum aspects of the emitted radiation, such as its angular momentum and entanglement. This issue is of great interest today, because it recently became possible to shape the quantum wavepacket of a single electron [32-35], imprinting it with OAM [32-34,36-38] or with other intriguing shapes [35,39].

## Conventional theory of the Čerenkov Effect

We begin by recalling the conventional theory of the ČR (Frank-Tamm [2,5,24,40]). A relativistic point charge is moving with velocity $v = \beta c$ inside a homogeneous medium with refractive index $n$, where $c$ is the speed of light. When the particle velocity is larger than the phase velocity of light in the medium, i.e. $\beta = v/c > 1/n$, radiation is emitted in a cone with the spread angle $\theta_{ph} = \theta_{\text{ČR}}^{conventional}$. The conventional ČR can be summarized by two equations:

$$\cos(\theta_{\text{ČR}}^{conventional}) = (\beta n)^{-1} \tag{1a}$$

$$\Gamma_\omega = \alpha\beta \sin^2(\theta_{\text{ČR}}^{conventional}) \ . \tag{1b}$$

Where $\alpha$ is the fine structure constant ($\approx 1/137$), and $\Gamma_\omega$ is the rate of photon emission **per unit frequency**. This rate depends on $\omega$ indirectly, with the only dependence being in the material dispersion by the substitution of $n = n(\omega)$. All the results we present in this work only depend on $\omega$ through this substitution, with no other dispersion correction (e.g., $n'(\omega)$ terms cancel out), hence we omit the $(\omega)$ notation and simply write $n$. Equations 1a&b are a direct outcome of Maxwell's equations with the electric and magnetic fields in the far-field limit, and a charge/current source being a moving delta function source (point particle). However, in the physical world point particles are never an exact description – all particles have some statistical distributions in their position and momentum or even quantum uncertainty. e.g., their momentum is characterized by some spread angle and some distribution of a finite width. These important characteristics are typically ignored without any significant impact on the results. But as we show below, they make an important difference for ČR.

Many papers have dealt with ČR in a quantum mechanical context (e.g., [25-29,41-43]), but always considered the particle as a single momentum state, analogous to a "plane-wave" in space. Under such an assumption, it was shown that one can re-derive the original equation for the rate, Eq.1b. However, a "plane-wave" description of a particle, being a single momentum state, is exactly a delta function in momentum space. Such a description is just as problematic as the classical description assuming a delta function in real space. Real charge carriers are neither plane-waves of a single momentum, nor singular point sources given by delta functions in space. It is this additional feature – of the ***coherent superposition of plane-waves with different momentum directions***, whose interference makes up the quantum wavepacket describing the electron – that gives rise to the new effects involving the OAM of the electron and photon. Thereby, engineering the quantum wavepacket can greatly enhance the predicted effects and show additional effects that are unique to specially shaped quantum particles. A direct implication is that one can control properties of the ČR by shaping the wavepacket of the incoming electron or ebeam. Henceforth we call our particle "electron", even though the results are exactly the same for any charged spin-½ fermion (since we make no assumptions with respect to the mass of the particle). Moreover, the generalization for particles with any spin is straightforward, only changing the spinor-dependent terms in the result (Eq.7), as was discussed in a review back in 1957 [28].

**Introducing the formalism: the quantum approach to the Čerenkov Effect**

In QED, a wavepacket is a sum (or integral) over creation operators of single electrons, where each describes an electron of a single momentum state. We describe these single electron states in momentum space using cylindrical coordinates, so that every state is defined by $\left|p_i^{cyl}\right\rangle =$

$|E_i, s_i, \theta_i, l_i\rangle$, which contains its energy $E_i$, spin $s_i$, spread angle $\theta_i$, and OAM $l_i$, measured with respect to the cylinder axis ($z$). See Fig.1 for an illustration of the cylindrical states and the notations. The longitudinal and transverse momenta are $p_{iz} = \beta E_i \cos(\theta_i)/c$ and $p_{ir} = \beta E_i \sin(\theta_i)/c$ respectively (where $E_i = (1-\beta^2)^{-1/2} mc^2$). When presenting this state in real-space – it has a Bessel function-like profile – as recently shown in [44] as wavepacket solutions of the Dirac equation. Notice that the cylindrical states always have their OAM mixed with the spin angular momentum, as presented in [44], and in the supplementary information (SI) section I.

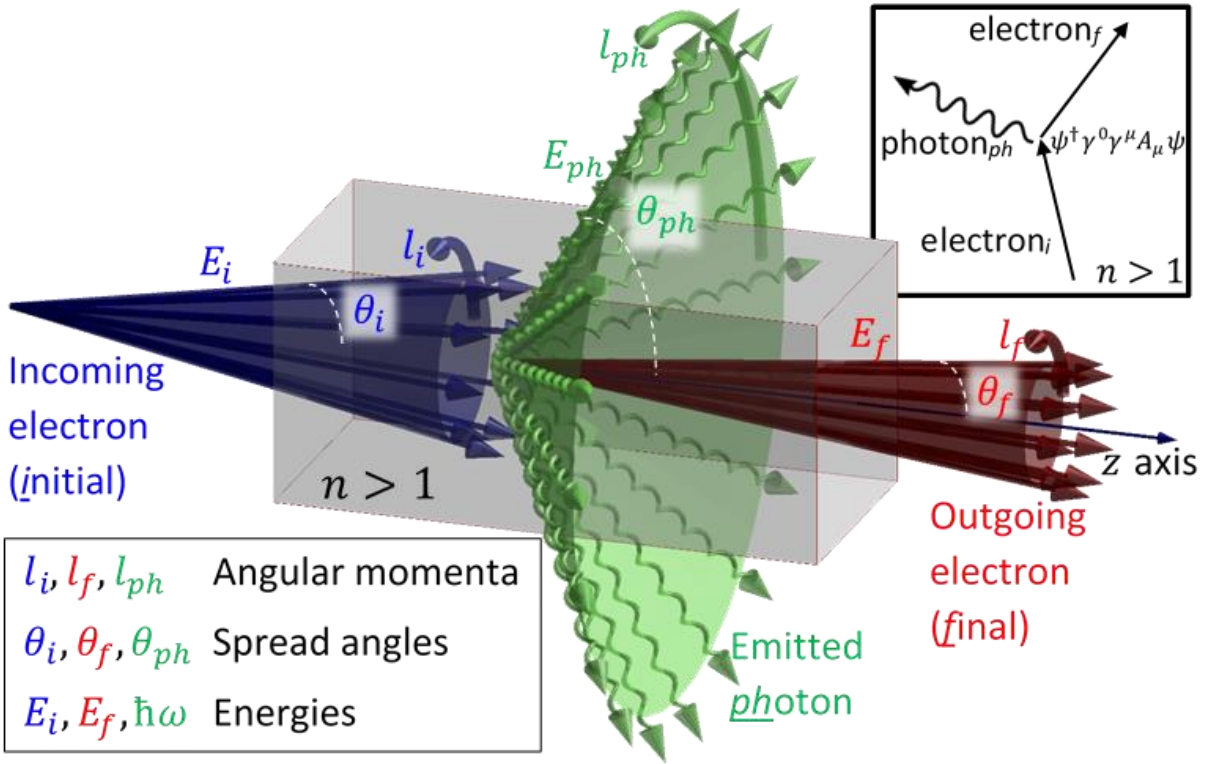

Figure 1: Illustration of the ČR process. The incoming (outgoing) electron is drawn in blue (red), and the emitted photon in green.

To develop the ČR in QED, we consider the spin-polarization term in the Dirac Hamiltonian $\psi^\dagger \gamma^0 \gamma^\mu A_\mu \psi$, describing the electron-photon interaction. Here, $\psi$ ($\psi^\dagger$) is the electron annihilation

(creation) operator, and $A_\mu$ is the electromagnetic field operator. We impose a final state of a single electron and a single photon, both described in cylindrical coordinates by $|p_f^{cyl}\rangle \otimes |k^{cyl}\rangle = |E_f, s_f, \theta_f, l_f\rangle \otimes |\hbar\omega, s_{ph}, \theta_{ph}, l_{ph}\rangle$ (Fig.1), forming a complete basis for the process. The electron parameters are as above, while the photon is characterized by its frequency $\omega$, polarization $s_{ph}$ (azimuthal/radial), emission angle $\theta_{ph}$, and OAM $l_{ph}$. We define the longitudinal and transverse photon wavenumbers $k_z = n\omega \cos(\theta_{ph})/c$ and $k_r = n\omega \sin(\theta_{ph})/c$ respectively (the dispersion relation of the photon is $\omega = |\vec{k}|nc$). In a medium where $n > 1$, this final state is created by a first-order interaction from an initial state of a single electron, and is ***the only possible first-order interaction***, hence it is the dominant effect. This first-order interaction does not occur in vacuum, because conservation of energy and momentum cannot be satisfied together.

The first step in describing the transition shown in Fig.1 ($p_i^{cyl} \to p_f^{cyl} + k^{cyl}$) is writing the bra-ket expression for the Hamiltonian density:

$$M^{density}_{p_i^{cyl} \to p_f^{cyl}+k^{cyl}}(t,r,\varphi,z) = \langle p_f^{cyl}\ electron,\ k^{cyl}\ photon | \underbrace{\psi^\dagger \gamma^0 \gamma^\mu \psi}_{\triangleq j^\mu(t,r,\varphi,z)} qA_\mu(t,r,\varphi,z) | p_i^{cyl}\ electron,\ 0 \rangle$$

$$= \underbrace{\left[\frac{q}{n\sqrt{\varepsilon_0}} \frac{\hbar c}{\sqrt{2\hbar\omega_k}} \frac{\sqrt{k_r}}{\sqrt{L_r L_z}}\right]}_{qA_\mu\ normalization} \underbrace{\frac{\sqrt{\frac{1}{\hbar}P_{fr}\frac{1}{\hbar}P_{ir}}}{L_r L_z}}_{j^\mu\ normalization} \exp\left(\underbrace{i\tfrac{1}{\hbar}(P_{iz}-P_{fz})z - ik_z z}_{conservation\ of\ on-axis\ momentum}\right) \exp\left(\underbrace{i\omega t - i\tfrac{1}{\hbar}(E_i - E_f)t}_{conservation\ of\ energy}\right). \quad (2)$$

$$\cdot \left[e^{il_i\varphi} J_{l_i}\left(\tfrac{1}{\hbar}P_{ir}r\right)\right]\left[e^{-il_f\varphi} J_{l_f}\left(\tfrac{1}{\hbar}P_{fr}r\right)\right]\left[e^{-i(l_{ph}-1)\varphi} J_{l_{ph}-1}(k_r r)\right] \underbrace{\left[\frac{P_{fz}(E_i+mc^2) - P_{iz}(E_f+mc^2)}{2\sqrt{E_i E_f}\sqrt{E_i+mc^2}\sqrt{E_f+mc^2}}\right]}_{spinor-polarization\ term}$$

Where the normalization of the current operator $j^\mu \triangleq \psi^\dagger \gamma^0 \gamma^\mu \psi$ uses the cylindrical radius $L_r$ and length $L_z$, which will be taken to infinity later (SI section II or [37,45,46]). In the normalization of the photon field operator $A^\mu$, $\varepsilon_0$ is the vacuum permittivity, and $\omega_k$ is the dispersion dependent normalization that later cancels out (see SI). The radial dependence is given in the form of Bessel functions, due to the cylindrical symmetry. Notice that Eq.2 is given for the special case of the electron flipping its spin and the photon having an azimuthal polarization. Each spin and choice of polarization results in a different spinor-polarization term, and in a change in the orders of the Bessel functions (SI sections I,II).

**Quantum derivation: the matrix element**

Integrating Eq.2 over space-time yields the matrix elements, or the amplitudes of the transition $p_i^{cyl} \to p_f^{cyl} + k^{cyl}$. We get delta functions for the conservation of energy, longitudinal momentum, and angular momentum. The latter shows a unique mixture of spin and OAM: any combination of outgoing electron and photon is allowed in Eq.2, as long as they satisfy a conservation of OAM, $l_{ph} + l_f \pm 1 = l_i$. The $\pm 1$ is because this transition involves a spin-flip, and will also appear in the case of spin-flip with radial polarization, but with different amplitudes. Accordingly, the transitions without spin-flip have a direct conservation of OAM $l_{ph} + l_f = l_i$. The full description is in the SI. The amplitudes of the ČR transitions are also affected by the OAM, since the electron and photon OAM set the orders of the Bessel functions in Eq.2. Plotting the amplitudes, Fig.2a&b exhibit preferred "stripes" of high amplitude at certain angles of emission $\theta_{ph}$; these stripes depend on the OAM ($l_{ph} = 4,8$ for Figs.2a,b). To observe these features, one has to measure the OAM of the outgoing state (electron and photon) and distinguish between different outgoing channels of emission for the ČR. Since an infinite number of channels are possible simultaneously, the full outgoing state in the ČR process is a complex electron-photon state, correlated through their OAM energy and spread angle. However, if one only measures the radiation spectrum, these OAM features are averaged out and their distinct marks completely disappear.

Exactly quantifying the amplitudes requires solving a triple-Bessel integral over the cylindrical radius $r$, which was fortunately studied in the mathematical literature [47] providing us with a closed-form solution:

$$\int_0^\infty J_{l_i}(p_{ir}r/\hbar) J_{l_f}(p_{fr}r/\hbar) J_{l_{ph}+1=l_i-l_f}(k_r r) r\, dr = \frac{\cos(l_i \alpha_f - l_f \alpha_i)}{2\pi S_\Delta\left(\frac{1}{\hbar} p_{ir}, \frac{1}{\hbar} p_{fr}, k_r\right)}, \tag{3}$$

where $S_\Delta$ is the area of a triangle with sides of lengths $\frac{1}{\hbar}p_{ir} = \frac{1}{\hbar}p_i \sin(\theta_i)$, $\frac{1}{\hbar}p_{fr} = \frac{1}{\hbar}p_f \sin(\theta_f)$, and $k_r = n\omega \sin(\theta_{ph})/c$, where $\alpha_i$, $\alpha_f$, and $\alpha_{ph}$ are the angles opposite the three sides. If a triangle cannot be made, then the integral is zero, which gives another selection rule (though not a simple one) for the possible radiation emission. When combined with the conservation of energy, momentum, and angular momentum, this selection rule defines a finite regime shown in Fig.2. Most importantly, the amplitude always diverges on the edge of the regime (blue/red/green dashed lines in Figs.2a&b), since this is where the triangle area $S_\Delta$ goes to zero. The integral of Eq.3 recently appeared in studies of scalar Compton backscattering, where these selection rules were discussed in the spinless case [45,46], along with the implications of finite integration range [46] and the nature of the $S_\Delta^{-1}$ divergence (appendices of [45]). This shows that our calculations can describe additional physical phenomena such as the Compton Effect from electrons or other fermions (since we solve the full vector case – i.e., taking into account the spin).

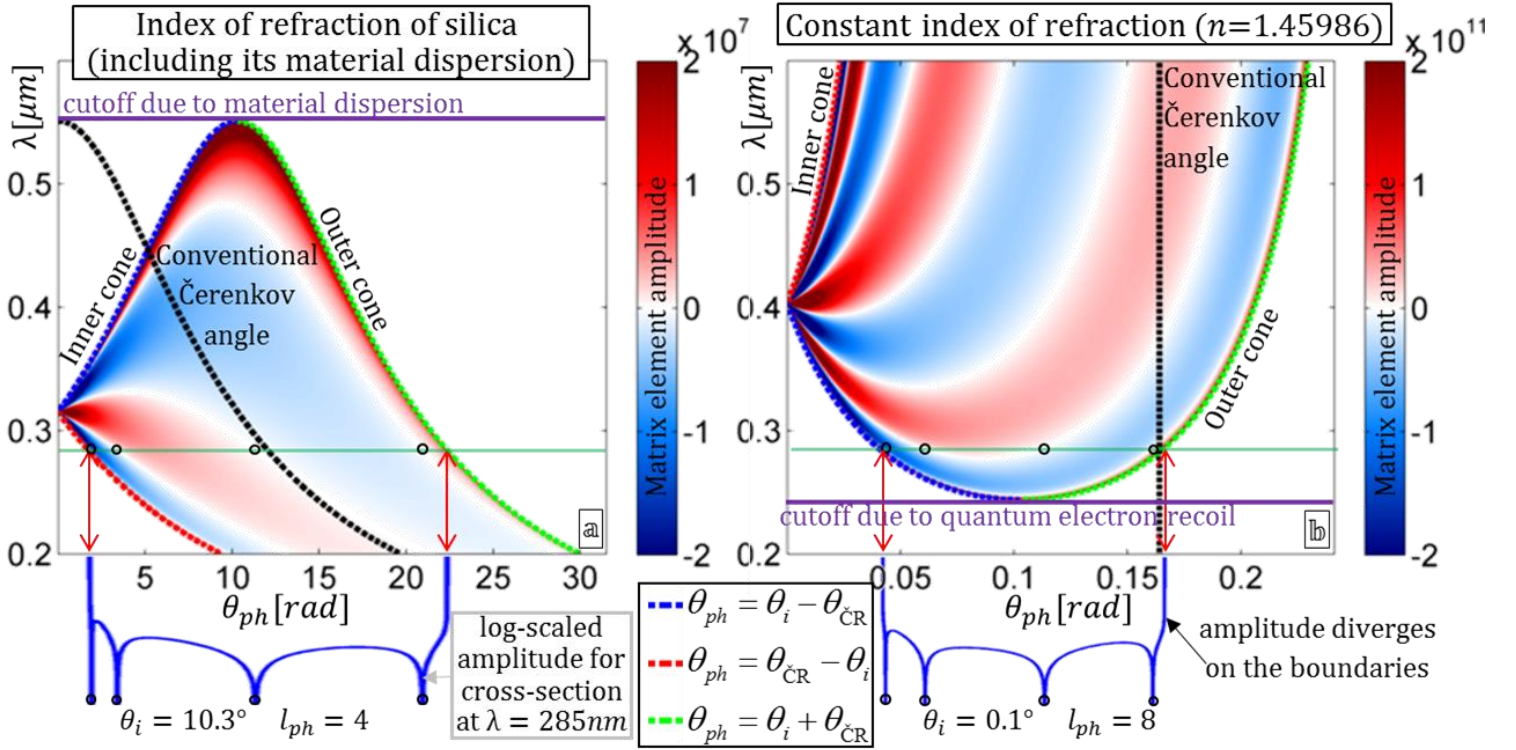

Figure 2: Matrix elements amplitudes for the ČR process, as a function of the photon wavelength $\lambda$ and emission spread angle $\theta_{ph}$. The colormap shows the spatial part of the matrix element (Eq.3) that vanishes outside of the permitted zone, bounded by the blue/red/green dashed curves. Along these curves the amplitude diverges; thus, we use a saturated color scale, with darker colors corresponding to higher transition amplitudes. The divergences are highlighted by the cross-section plots below the maps also showing the nodal lines of zero amplitude (marked by black circles) between high amplitude "stripes". These distinct "stripes" depend on the OAM ($l_{ph} = 4, 8$ in (a,b)), showing the coupling between the OAM and preferred emission angles inside the permitted zone (which is independent of angular momentum). The black dashed curve denotes the angle of the conventional ČR (Eq.1a), changing with $\lambda$ in (a) due to the silica dispersion. The spectral cutoffs are marked by solid purple lines – beyond these wavelengths no photons are emitted. At the points tangential to this line $\theta_{\check{C}R} = 0$ hence $\theta_{ph} = \theta_i$ (equals 10.3° in (a) or 0.1° in (b)). All figures have $\beta = 0.685$, and refractive index of silica including its material dispersion (a) or constant $n = 1.45986$ (b).

Due to the divergence along the edges, the photon emission primarily occurs along discrete angles $\theta_{ph} = |\theta_i \pm \theta_{\check{C}R}|$ marked by the blue/red/green dashed lines. The formula for $\theta_{\check{C}R}$ in the quantum derivation (generalizing Eq.1a) is still independent of $\theta_i$, hence it is most easily found for the limit case of $\theta_i = 0$, where our incoming electron state $|p_i^{cyl}\rangle$ reduces to a plane-wave (a single momentum state $p_i = \beta E_i / c$), and the radiation spread angle is just $\theta_{ph} = \theta_{\check{C}R}$ (black

dashed line). Then, $\theta_{\text{ČR}}$ is found analytically from elementary conservation laws, as was first shown by Sokolov [41], and later by Cox [48] and many others [26-28,42,43]:

$$\theta_{\text{ČR}} = \arccos\left(\frac{1}{\beta n} + \frac{\omega}{\omega_C}\frac{\sqrt{1-\beta^2}}{\beta}\frac{n^2-1}{2n}\right), \tag{4}$$

where $\omega_C = mc^2/\hbar$ is the Compton frequency, given by the rest energy of the electron divided by the reduced Planck Constant. To avoid confusion, notice that for incoming $|p_i^{cyl}\rangle$, we use $\theta_{\text{ČR}}$ only as a convenient notation, and not as the angle of photon emission $\theta_{ph}$ (compare black vs. blue/red/green dashed lines). This significant difference has an important physical implication: the $|p_i^{cyl}\rangle$ electron primarily has **two** allowed angles of ČR emission for each $\omega$ ($\theta_{ph} = |\theta_i \pm \theta_{\text{ČR}}|$), resulting in a double Čerenkov cone, which we name the inner ($\theta_{ph} = |\theta_i - \theta_{\text{ČR}}|$) and outer ($\theta_{ph} = \theta_i + \theta_{\text{ČR}}$) cones. Interestingly, for any $\theta_i > 0$ and high enough electron velocity, the outer cone becomes a ***backward cone*** ($\theta_{ph} > 90°$), which is supposed to be impossible in ordinary materials [8,9], under the conventionally used Čerenkov theory. These two angles can be understood by noting that the state $p_i^{cyl}$ is a superposition of plane-waves with their momenta at an angle $\theta_i$, each emitting ČR at a relative angle $\theta_{\text{ČR}}$. More generally, the cone splitting originates from the *shape* of the charge distribution and is not necessarily unique to the quantum ČR. Namely, it can also occur for classical charge distributions having a well-defined spread angle $\theta_i$, as well as in analogues of the Čerenkov Effect in other areas of physics. In either case, the cone splitting we show here is uniquely tied to the shape of the incoming electron wavepacket (or the charge distribution in a classical electron beam), and is independent of material properties that can cause other kinds of cone splittings [49].

Another important consequence highlighted in Fig.2b is marked by a solid purple line that bounds the range of emission by a spectral cutoff. This is in contradiction to Eqs.1a&b that have no bound and are even frequency-independent. Namely, the conventional ČR is famously broad in spectrum, truncated only by the material dispersion (when $n = n(\omega)$ drops towards 1 and gets below $1/\beta$, as shown by the cutoff in Figs.2a,3a&c). In contradistinction with that, the quantum derivation yields a fundamental frequency cutoff that exist irrespective of material dispersion (as shown in the by the cutoff in Figs.2b,3b&d):

$$\omega_{cutoff} = 2\omega_C \frac{\beta n - 1}{(n^2 - 1)\sqrt{1-\beta^2}}, \quad (5)$$

The fact that a frequency cutoff appears at some extreme frequency ($\omega_C$ is in the gamma-rays) is expected – a trivial reason being that a charged particle cannot create a photon carrying more energy than it originally had. This was indeed mentioned in some papers as an extreme limit for the ČR (see e.g. [41]). However, it was always thought to be of no practical interest, since *"as shown by Tamm, the spectrum must be cut off at a smaller frequency"* ([41]). That is, $\omega_{cutoff}$ was believed to be much larger than the conventional cutoff that comes from the material dispersion ($\omega$ for which $n(\omega)$ drops below $1/\beta$ [24]). Indeed, the quantum correction to ČR at frequencies much lower than $\omega_{cutoff}$ is obviously insignificant. This logic led Landau and Ginzburg to reason that the quantum corrections must be immaterial in the optical region, as stated explicitly by Ginzburg in 1996 [43], when referring to his PhD paper from 1940 [25]: *"In the optical region, the only one where applications of the VC effect are normally feasible, the ratio ħω/mc² ~ 10⁻⁵ [very small] even for electrons, i.e. quantum corrections are immaterial.    In 1940, L D Landau told about my work [25] stated that it was of no interest (see Ref. [50] p. 380). It follows from the above, that he was fully justified in drawing this conclusion, and his comment hit the mark as was usual with his criticism"*.

Evidently, Ginzburg and Landau believed that the quantum correction to the ČR effect has no implication in the optical spectral region. However, a particularly interesting implication of Eq. 5, which as far as we know was not noted earlier, ***does concern the optical region after all***: near the Čerenkov threshold the velocity satisfies $\beta \approx 1/n$, and thus $\omega_{cutoff}$ shifts to very low frequencies. Equation 5 shows that the cutoff is linear in $\beta - 1/n$, with a very large slope ($\sim mc^2/\hbar$). Thus, taking a charged particle with a velocity arbitrarily close to the Čerenkov threshold would shift the frequency cutoff all the way to zero. In practice, this is only possible up to the experimental precision in setting the velocity of the charged particle, or equivalently, its energy variance. Fortunately, variances in the electron energies in modern electron guns are already low enough. For example, transmission electron microscopes (TEM- typical energy 100-200KeV) can work with a variance in the electron energy lower than 1eV, which brings the cutoff to the optical frequencies (2eV photon = wavelength of 620nm). Furthermore, variances in the value of the refractive index $n$ in high quality glasses are even smaller ($\sim 10^{-6}$ with three terms in the Sellmeier equation). Moreover, as we show below, even with a larger variances, the predicted effect can still be observed, because ČR gets a new frequency-dependent correction.

**Quantum derivation: the rate of emission**

So far we have found the amplitude of the general $p_i^{cyl} \to p_f^{cyl} + k^{cyl}$ transition. Below, we are interested in the rate of ***photon emission***, insensitive to the momentum of the outgoing electron, which is the simplest observable quantity (being what a spectrometer would measure). We integrate over all 3D momentum space of the outgoing electron, finding the rate of photon emission for any value of the outgoing photon momentum. When also integrating over the (solid) angles of radiation, we obtain the spectrum of ČR, or the rate of photon emission per unit frequency $\Gamma_\omega$, as shown in Eq.1b. Importantly, these rates of ČR emission are independent of the OAM, and in general are insensitive to any phase information of the incoming wavepacket. Consequently, all the results henceforth also describe incoherent wavepackets and classical electron distribution.

The summation over the outgoing electron momenta eliminates the delta functions in the matrix elements, and removes the dependence on a finite cylindrical box in the limit of $L_r, L_z \to \infty$. The elaborate calculations needed to solve the integrals are all done analytically (SI sections IV,V), for any functional dependence of the refractive index $n = n(\omega)$. Eventually, we find the rate of photon emission $\Gamma_\omega$ in four distinct cases that differ by the azimuthal/radial polarization and the spin flip/non-flip. The case of azimuthal emitted photon and spin-flip is given by: (the other three cases have similar expressions presented and explained in the SI sections V,VI)

$$\Gamma_{\omega, azimuthal, \uparrow\downarrow} = \frac{\alpha}{4\beta}\left(\frac{\hbar\omega}{E_i}\right)^2 \left\{\cos^2(\theta_i)\frac{\left(\beta E_i - n(E_i + mc^2)\cos(\theta_{CR})\right)^2}{(E_i + mc^2)(E_f + mc^2)} + \frac{1}{2}n^2\sin^2(\theta_i)\sin^2(\theta_{CR})\frac{E_i + mc^2}{E_f + mc^2}\right\} \quad (6)$$

Studying Eq.6 near any frequency cutoff (either the quantum cutoff $\omega_{cutoff}$ or any cutoff created by material dispersion) shows a spectral ***discontinuity***, because on one side the emission vanishes when $\theta_{ČR} = 0$, yet on the other side the left term in Eq.6 remains finite. This is in contradiction to the inevitable zero rate found beyond the cutoff, implying a discontinuity exactly at the cutoff, as is shown in Figs.3a&b. Notice that conventional ČR has no such discontinuity whatsoever (Eq.1b), as also shown by the continuous curves in Figs.3c&d. Observing this effect would be challenging because it is much smaller than the conventional ČR rate, scaling like $(\hbar\omega/E_i)^2$. However, recent progress in spin-polarized ebeams, and especially spin-polarized transmission electron microscope [51], might create the opportunity to observe this discontinuity in the energy loss of electrons by filtering only the cases of electron spin-flip.

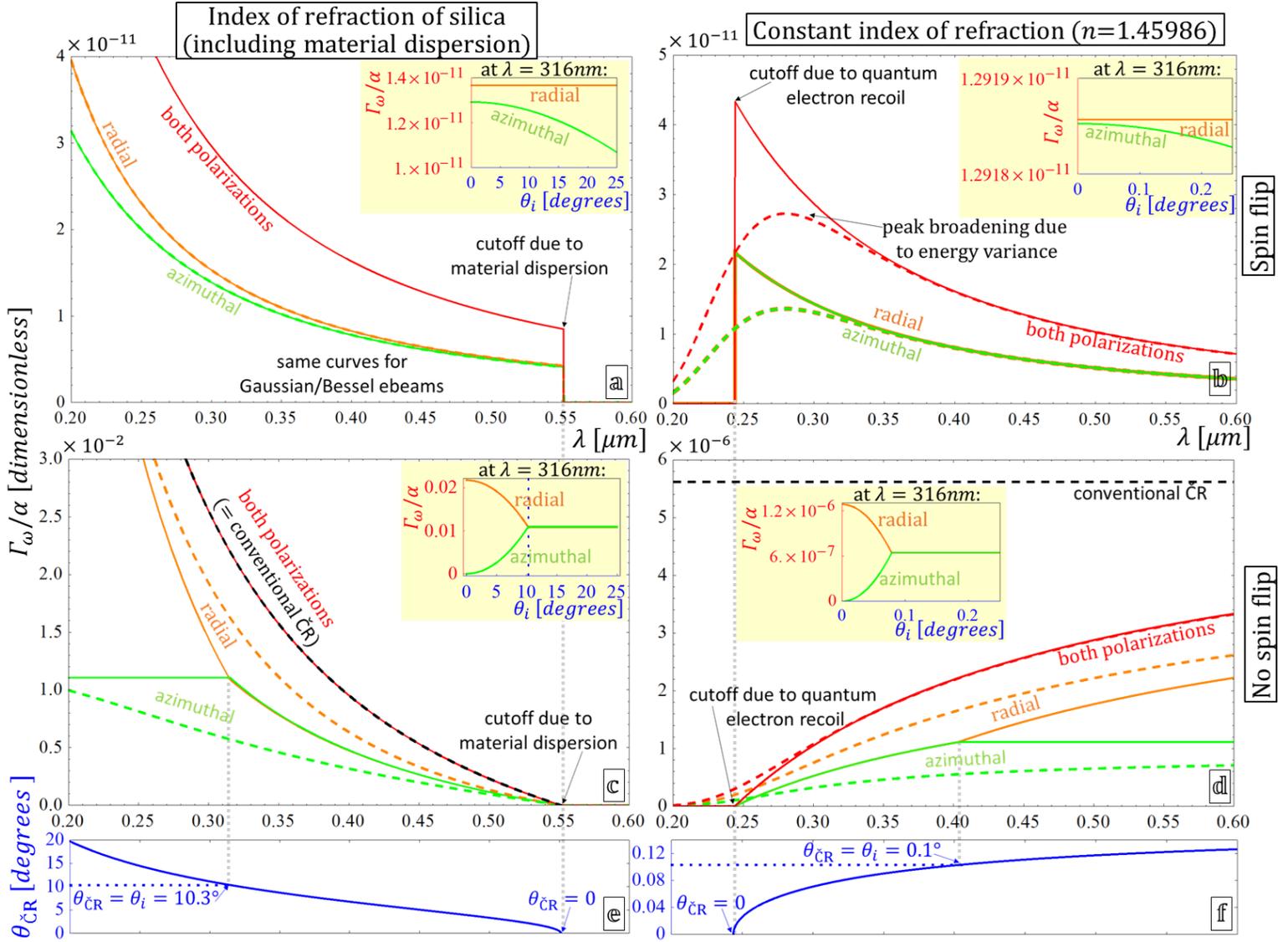

Figure 3: Deviations between classical and quantum ČR theories – conventional vs. QED photon emission rates, demonstrated with Bessel ebeams (solid curves) and Gaussian ebeams (dashed curves). Dashed black: conventional ČR result, according to Frank-Tamm (Eq.1b). Green and orange: quantum ČR for emission into azimuthal and radial polarizations (red curve is for their sum). Blue: the Čerenkov angle (in (e,f)), showing a spectral cutoffs where it becomes zero, and spectral kinks where it crosses $\theta_i$, both marked by gray dotted lines. Unlike the conventional Čerenkov theory, the quantum theory predicts ČR that involves electron spin flip (panels (a,b)) having discontinuities at both cutoffs. The dominant part of the ČR emission (no spin flip, panels (c,d)) matches the conventional theory in (c), but deviates from it near $\omega_{cutoff}$ in (d). The kinks (abrupt change in slope), which are also emphasized in the insets, are a unique feature of the Bessel ebeam that is smoothened-out in Gaussian ebeams – compare solid and dashed curves in (c,d). For the current choice of parameters (a,b) are practically agnostic to $\theta_i$. Panels (a,c) present ČR in silica for Bessel ebeam with $\theta_i = 10.3°$ or Gaussian with spread FWHM of $10.3°$, both showing a "trivial" spectral cutoff at $\lambda = 550nm$ (occurs since $n$ drops below $1/\beta$). Panels (b,d) present ČR assuming a constant $n = 1.45986$ for Bessel ebeam with $\theta_i = 0.1°$ or Gaussian with spread FWHM of $0.1°$, both showing a new spectral cutoff at $\lambda = 244nm$ (occurs due to the quantum correction to ČR from electron recoil, as shown by $\omega_{cutoff}$ in Eq.5). Different choice of $n$ causes $\omega_{cutoff}$ to shift, yet its exact value only matters at the proximity of the cutoff: no need for a constant $n$ over a wide spectrum. All ebeams have $\beta = 0.685$ (~190keV); while the Bessel ebeams are mono-energetic, the Gaussians have energy uncertainty of $\Delta E = 0.5eV$. The predicted effects are not special for the current choice of parameters.

The rate $\Gamma_\omega$ shows another important features not predicted by the conventional ČR theory for cases of no spin-flip when near the quantum cutoff $\omega_{cutoff}$ (Fig.3d). This deviation can even be observed without measuring the spin at all, and has a much larger amplitude than the previous (spin-dependent) case. Summing over all possible outgoing spin and polarization states we obtain:

$$\Gamma_{\omega,total} = \alpha\beta \sin^2(\theta_{\check{C}R}) + \frac{\alpha}{\beta}\left(\frac{\hbar\omega}{E_i}\right)^2 \frac{n^2-1}{2} \qquad (7)$$

Equation 7, when taken together with Eq.5, shows the quantum generalization of Eqs.1a&b. These conform to the conventional result when assuming $\hbar\omega \ll E_i$. The important difference we find between them is a deviation near $\omega \approx \omega_{cutoff}$ as shown in Fig.3d: comparing the dashed black curve to the green/orange/red solid curves – shows a pronounced difference for wavelength shorter than the cutoff, between a finite rate of Eqs.1a&b and the zero rate of Eqs.5&7. Of course, observing this deviation requires a careful tuning of $n$, at least for a small window of wavelengths around the cutoff. Recent advancement in fabrication methods in nanophotonics, and especially metamaterials and photonic crystals, are exactly up to this task, allowing exact tuning of $n(\omega)$ in specific windows of parameters [15].

**Quantum corrections occurring even without specific shaping: Gaussian ebeams**

All the effects above are found for a shaped incoming electron having a well-defined spread $\theta_i$, describing a single cylindrical momentum state $|p_i^{cyl}\rangle$ or a Bessel ebeam. Will the effects remain for ebeams or wavepackets that are not specially shaped? The total rate in Eq.7 is very general as $\Gamma_{\omega,total}$ is independent of the spread $\theta_i$; it also occurs irrespective of the electron being a coherent wavepacket or an incoherent beam – as expected from the incoherent summation in the calculation (SI section IV). This means that the total rate of emission is the same for any shaped ebeam (e.g., Gaussian, Bessel, or plane-wave), which is probably why no previous experiment has observed deviations from the conventional theory (as the primary observable in such experiments was the emission rate). Nevertheless, the total rate is the only result that does not depend on $\theta_i$ – when measuring either the emission direction, polarization, or any property of the electron, we have shown strong $\theta_i$ dependence (e.g., Eqs.2,3,6 and Figs.2&3). In order to generalize the calculations to arbitrary shaped ebeams one can directly integrate the current expressions over $\theta_i$ or $E_i$ with proper weight functions. As examples, we calculate the ČR emission rates for Gaussian ebeams (dashed curves in Fig.3) that have variance in both their angular spread $\theta_i$ (variance of 10.3° in Figs.a&c, or 0.1° in Figs.b&d) and energy $E_i$ ($\Delta E_i = 0.5 eV$, a conservative TEM value; modern microscopes used for EELS reach variances smaller by more than an order of magnitude). The important fact is that the effects due to the quantum nature of the particle do survive: the discontinuities at the cutoffs and the resulting deviations from the conventional theory are also observable for these Gaussian ebeams. In particular, the discontinuity in Figs.3a is practically unaffected (dashed and solid curves coincide), while the discontinuity in Fig.3b is partially smoothened (spectral broadening), yet there is still a noticeable peak. Figure 3d has a similar broadening, nonetheless, the deviation between the quantum and the conventional results (dashed red vs. dashed black curves) remains for even larger energy variances. It proves that the deviations arising from the quantum calculation are not exclusive for the $|p_i^{cyl}\rangle$ states – they occur for any ebeam that is a real physical entity. Importantly, the results shown in the dashed curves in Fig.3 are not specific for Gaussians: they appear for any realistic wavepacket we tried. As such, they should appear in nature without intentional shaping of the wavepacket, and should be observable in the spectrum provided a small enough variance in the particle energy.

Other sources of spectral broadening are the limited penetration depth of the electron into the medium, optical losses in the medium, and the finite volume of this medium [52]. However, these effects are known from conventional ČR; here they usually present a smaller broadening compared with the electron energy variance. For an exact expression taking into account the finite system and finite interaction size, as well as the possibility of material losses, one can modify the calculation in the SI by replacing the delta functions with Lorentzians or sincs and compute the integrals numerically [53].

**Conclusions and discussion**

To conclude, we would like to discuss the general insight resulting from this work: a realistic particle is never in just a single momentum state, therefore, ***any scattering process should involve similar quantum corrections that follow from the particle wave structure***. The corrections can be significant, as long as the length scale of the process is not exceedingly larger than the size of the wavefunctions of the particles involved. This happens in other electron-photon interactions (e.g. Compton effect), but is not limited to these, and can in principle happen in any scattering process in particle physics when there is enough uncertainty in the positions of the particles. Applications of our work include future Čerenkov detectors that reveal more information about a particle by also measuring the radiation spectrum, spread, or polarization, instead of just counting photons. Furthermore, the coupling between the OAM of the final electron and the photon suggests means of creating an entangled pair of a photon and a free-electron, as well as a new approach for the creation of entangled photons, controlled by the shaping of the incoming electron.

**Acknowledgements**

We would like to thank Prof. Levi Schächter, Prof. Robert L. Jaffe, Prof. Igor Ivanov, and Prof. Avi Gover for fruitful discussions. We also thank Prof. Francois Ziegler and Prof. Carlo Beenakker for valuable suggestions regarding the triple-Bessel integral.

This research was supported by the Transformative Science Program of the Binational USA-Israel Science Foundation (BSF) and by the Israeli Excellence Center "Circle of Light". Also, the work was supported in part by the U.S. Army Research Laboratory and the U.S. Army Research Office through the Institute for Soldier Nanotechnologies, under contract number W911NF-13-D-0001. The research of I.K. was partially supported by the Seventh Framework Programme of the European Research Council (FP7- Marie Curie IOF) under grant agreement n° 328853 – MC--BSiCS.

**Figure 1**

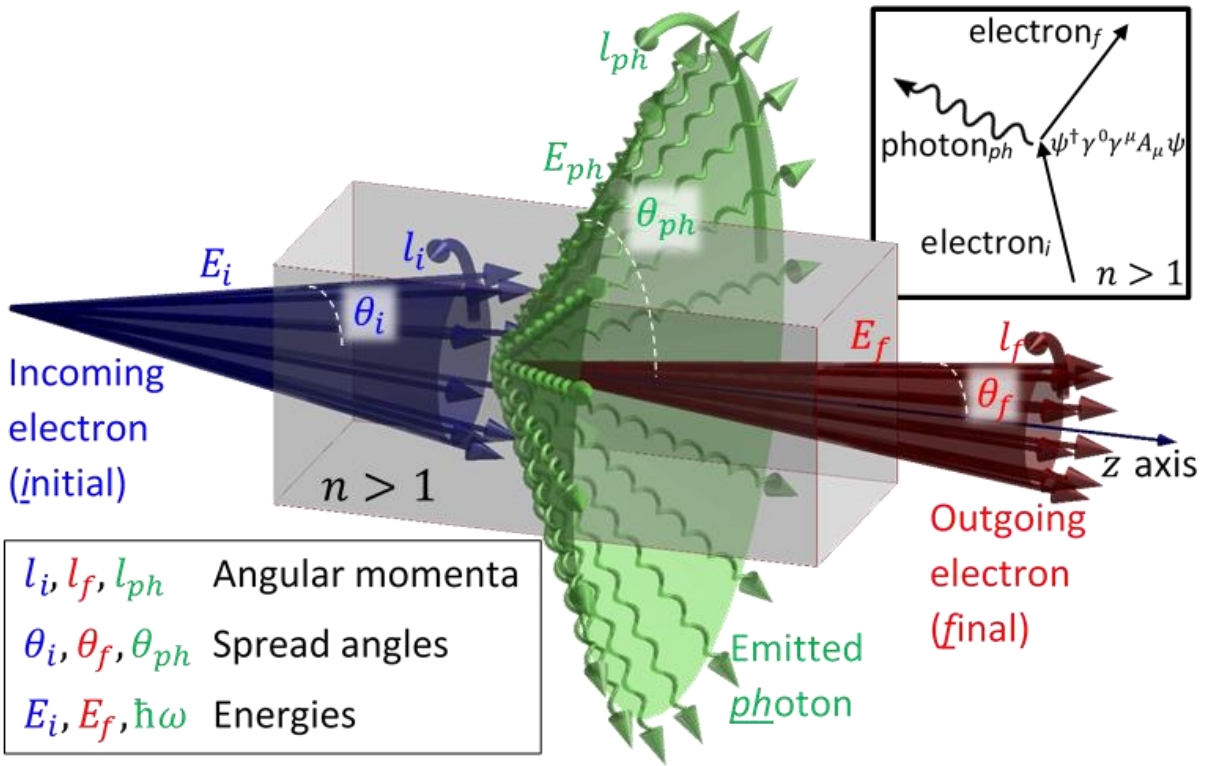

Figure 1: Illustration of the ČR process. The incoming (outgoing) electron is drawn in blue (red), and the emitted photon in green.

**Figure 2**

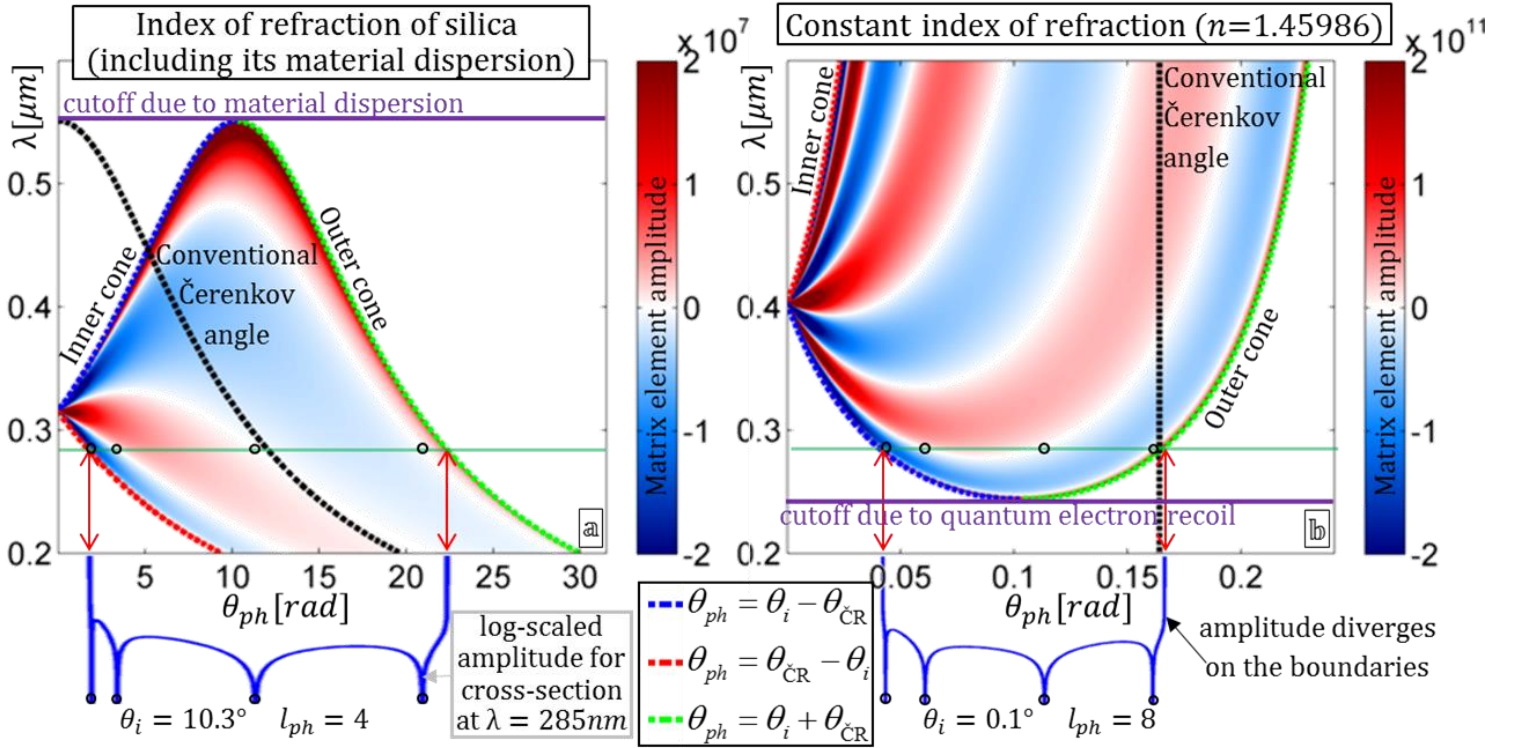

Figure 2: Matrix elements amplitudes for the ČR process, as a function of the photon wavelength $\lambda$ and emission spread angle $\theta_{ph}$. The colormap shows the spatial part of the matrix element (Eq.3) that vanishes outside of the permitted zone, bounded by the blue/red/green dashed curves. Along these curves the amplitude diverges; thus, we use a saturated color scale, with darker colors corresponding to higher transition amplitudes. The divergences are highlighted by the cross-section plots below the maps also showing the nodal lines of zero amplitude (marked by black circles) between high amplitude "stripes". These distinct "stripes" depend on the OAM ($l_{ph} = 4,8$ in (a,b)), showing the coupling between the OAM and preferred emission angles inside the permitted zone (which is independent of angular momentum). The black dashed curve denotes the angle of the conventional ČR (Eq.1a), changing with $\lambda$ in (a) due to the silica dispersion. The spectral cutoffs are marked by solid purple lines – beyond these wavelengths no photons are emitted. At the points tangential to this line $\theta_{\text{ČR}} = 0$ hence $\theta_{ph} = \theta_i$ (equals 10.3° in (a) or 0.1° in (b)). All figures have $\beta = 0.685$, and refractive index of silica including its material dispersion (a) or constant $n = 1.45986$ (b).

**Figure 3**

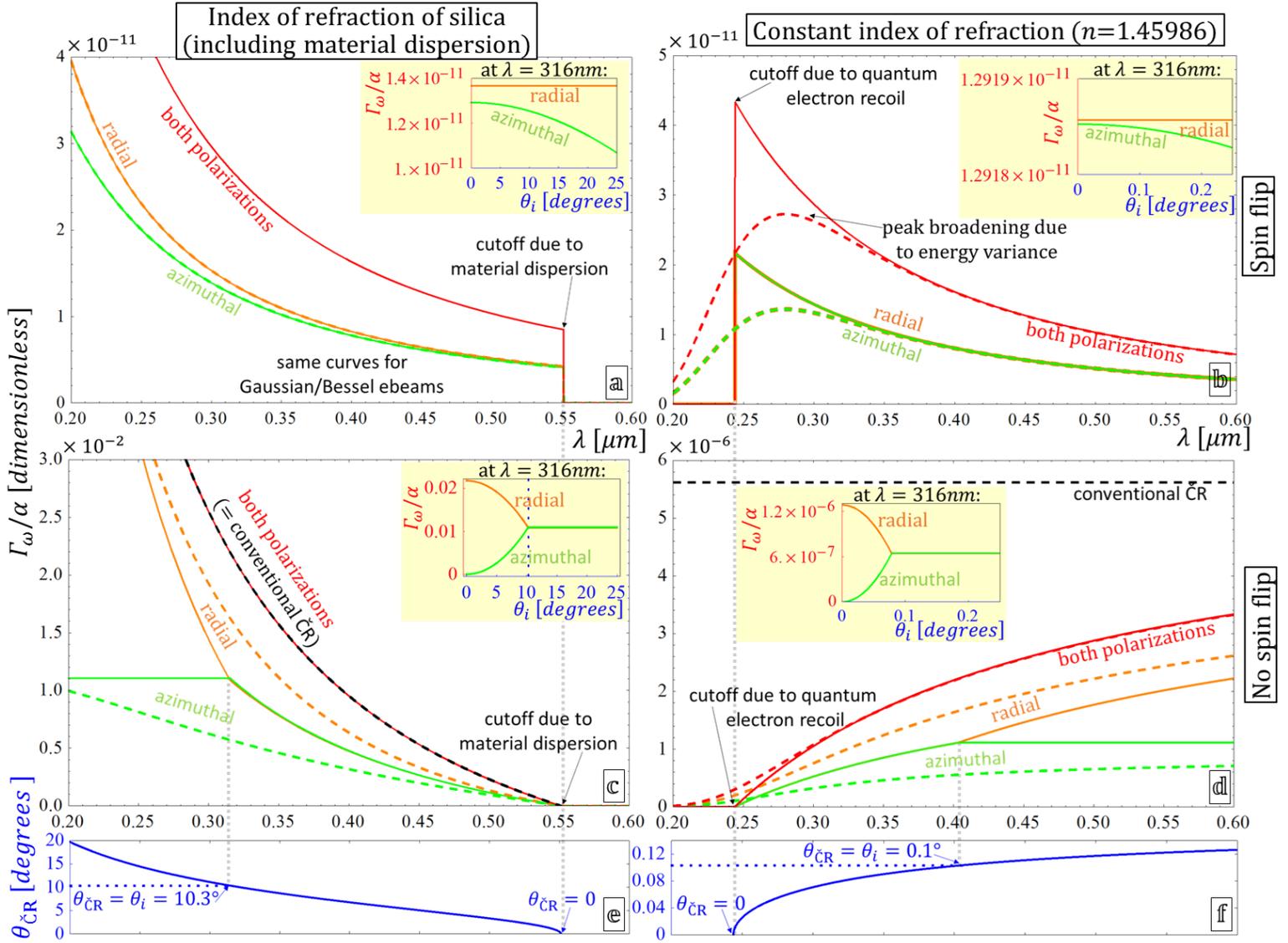

Figure 3: Deviations between classical and quantum ČR theories – conventional vs. QED photon emission rates, demonstrated with Bessel ebeams (solid curves) and Gaussian ebeams (dashed curves). Dashed black: conventional ČR result, according to Frank-Tamm (Eq.1b). Green and orange: quantum ČR for emission into azimuthal and radial polarizations (red curve is for their sum). Blue: the Čerenkov angle (in (e,f)), showing a spectral cutoffs where it becomes zero, and spectral kinks where it crosses $\theta_i$, both marked by gray dotted lines. Unlike the conventional Čerenkov theory, the quantum theory predicts ČR that involves electron spin flip (panels (a,b)) having discontinuities at both cutoffs. The dominant part of the ČR emission (no spin flip, panels (c,d)) matches the conventional theory in (c), but deviates from it near $\omega_{cutoff}$ in (d). The kinks (abrupt change in slope), which are also emphasized in the insets, are a unique feature of the Bessel ebeam that is smoothened-out in Gaussian ebeams – compare solid and dashed curves in (c,d). For the current choice of parameters (a,b) are practically agnostic to $\theta_i$. Panels (a,c) present ČR in silica for Bessel ebeam with $\theta_i = 10.3°$ or Gaussian with spread FWHM of 10.3°, both showing a "trivial" spectral cutoff at

$\lambda = 550nm$ (occurs since $n$ drops below $1/\beta$). Panels (b,d) present ČR assuming a constant $n = 1.45986$ for Bessel ebeam with $\theta_i = 0.1°$ or Gaussian with spread FWHM of $0.1°$, both showing a new spectral cutoff at $\lambda = 244nm$ (occurs due to the quantum correction to ČR from electron recoil, as shown by $\omega_{cutoff}$ in Eq.5). Different choice of $n$ causes $\omega_{cutoff}$ to shift, yet its exact value only matters at the proximity of the cutoff: no need for a constant $n$ over a wide spectrum. All ebeams have $\beta = 0.685$ (~$190keV$); while the Bessel ebeams are mono-energetic, the Gaussians have energy uncertainty of $\Delta E = 0.5eV$. The predicted effects are not special for the current choice of parameters.